\documentclass{PoS}

\title{The INTEGRAL/IBIS Complete Sample of Type 1 AGN}

\ShortTitle{The INTEGRAL/IBIS Complete Sample of Type 1 AGN}

\author{\speaker{M. Molina}$^{a}$, L. Bassani$^{a}$, A. Malizia$^{a}$, A. Bazzano$^{b}$, P. Ubertini$^{b}$ and
A.J. Bird$^{c}$\\
\llap{$^{a}$} INAF/IASF Bologna, Italy\\
\llap{$^{b}$} INAF -- IAPS Rome, Italy\\
\llap{$^{c}$} School of Physics and Astronomy, University of Southampton, UK\\ 
        E-mail: \email{molina@iasfbo.inaf.it}}
        
\abstract{The determination of the broad (0.1--100\,keV) spectra of active galaxies is crucial for 
understanding and discriminating among emission models, for estimating the properties of the 
Comptonising/reflecting region around the central black hole and for obtaining a firm description of 
the contribution of AGN to the Cosmic X-ray Background. 
Although broad-band X-ray measurements of AGN have been made in the past, these did not generally 
pertain to a complete sample of sources. Since few years, we have started a systematic analysis of the 
0.1--100\,keV spectra of a complete sample of AGN selected in the hard X-ray band (20--40\,keV) using 
low energy data (not always of good quality) from a set of operating X-ray telescopes. Thanks to data 
obtained through an {\it XMM-Newton} Large Programme, we have now high quality 0.1--10\,keV data for 
all sources in the sample; these combined with high energy observations 
from {\it INTEGRAL/IBIS} and {\it Swift/BAT} will allow us to study the spectral 
properties of this complete sample. Here in particular, we report the progress made on type 1 AGN, 
focusing in particular on the continuum and its high energy cut-off, the reflection fraction, 
the absorption properties and the presence of soft excesses and warm absorbers in our sources. More 
specifically we discuss the broad-band properties of 4 sources IGR J00333+6122, Swift J0917.2--6221, 
GRS 1734--292 and NGC6814, which can be considered as the most representative objects of our sample.}

\FullConference{An INTEGRAL view of the high-energy sky (the first 10 years) - 
9th INTEGRAL Workshop and celebration of the 10th anniversary of the launch,\\
		October 15-19, 2012\\
		Bibliotheque Nationale de France, Paris, France}

\begin{document}

\section{Introduction}
The determination of the slope of the continuum emission of AGN and its high energy cut-off
is essential for spectral modeling of AGN, since these two parameters 
are deeply linked to the physical characteristics of the Comptonising region around the central 
nucleus. Variations to this basic model depend on the energy distribution of the electrons 
and their location in relation to the accretion disc. Measuring both the primary continuum and its 
cut-off energy is therefore crucial for understanding models and discriminating between them. 
Another important ingredient in AGN spectral study is the reflection of the primary continuum which 
generates two features: a neutral iron K$\alpha$ line (at 6.4\,keV in the local reference frame)
and a reflection hump peaking at E$\simeq$30\,keV \cite{magdziarz95}. Both are key ingredients to 
study the region around the central black hole and eventually to test unified models 
\cite{ricci11}.
The determination of the continuum and its high-energy cut-off, together 
with the covering fraction and the geometry of the cold dense gas responsible for the
reflection hump at around 30\,keV, is also essential in AGN synthesis models in order
to correctly reproduce the shape of the Cosmic X-ray Background (CXB). All this information
can be obtained by means of broad-band (0.1--100\,keV)
spectral studies of samples of AGN.

However, so far such studies focused on few bright objects and 
never pertained to a complete sample of sources. 
This situation has changed since the launch of hard X-ray telescopes such as {\it IBIS} on board 
{\it INTEGRAL} and {\it BAT} on board {\it Swift}. We have already started to analyse broad-band 
spectra of a complete sample of AGN, selected in the 20--40\,keV band, using X-ray data from various 
X-ray instruments \cite{molina09, panessa11, derosa12} and high energy
data points from {\it IBIS/BAT}. In the meantime, we have requested and 
obtained to observe all sources through an {\it XMM-Newton} Large Programme. Here we present results from 
this follow-up programme focused on type 1 AGN \cite{molina09}, with the main goal of studying, thanks 
to high quality broad-band spectra, the continuum and its high energy cut-off, the reflection 
fraction, the absorption properties and the presence of soft excesses and warm absorbers in our 
sources.

\section{Sample and results}
The {\it INTEGRAL/IBIS} complete sample of type 1 AGN has been presented and analysed in 
\cite{molina09}. In this previous work broad-band spectra of the Seyfert 1--1.5 
listed in the {\it INTEGRAL} complete sample have been presented and discussed, 
mainly employing data from {\it XMM-Newton}, {\it Swift-XRT}, {\it Chandra} and {\it ASCA}.
The broad-band spectra of type 1 AGN in the {\it INTEGRAL/IBIS} complete sample were fitted employing 
an exponentially cut-off power-law, absorbed by photoelectric absorption and 
reflected from neutral material, where the reflection component 
is described by the parameter R = $\Omega$/2$\pi$. The sources in the sample are found to have 
absorbing column densities which are generally small or absent, 
except in those cases where complex absorption is required. The average power-law photon index is 
$\sim$1.7, a value which is flatter than the generally accepted canonical one of 1.9. The high-energy 
cut-off is found to be in the range 50--150\,keV ($<$E$_{\rm cut}$$>$ = 110\,keV). The average 
reflection fraction is 1.5, while the measured iron lines, in general, are found to be narrow, with 
only 15\% of the sample requiring a broad iron line. The line equivalent widths are below or around 
100\,eV, in general agreement with the measured reflection fractions.
Thanks to dedicated {\it XMM} observations, we have now high quality X-ray spectra of all the sources 
in the complete sample, which combined with {\it INTEGRAL/IBIS} and {\it Swift/BAT} data 
will allow us to study the broad-band spectral properties of all sources in more detail than before.
The main goal of this work is an in-depth study of the 0.5--100\,keV continuum, i.e. its slope and 
high energy cut-off, the study of the reflection fraction and of the absorption properties.
In the following, we present some preliminary results on few of the sources for which new {\it XMM}
data are now available.

\begin{small}
\begin{figure}
\centering
\includegraphics[angle=-90, width=0.47\linewidth]{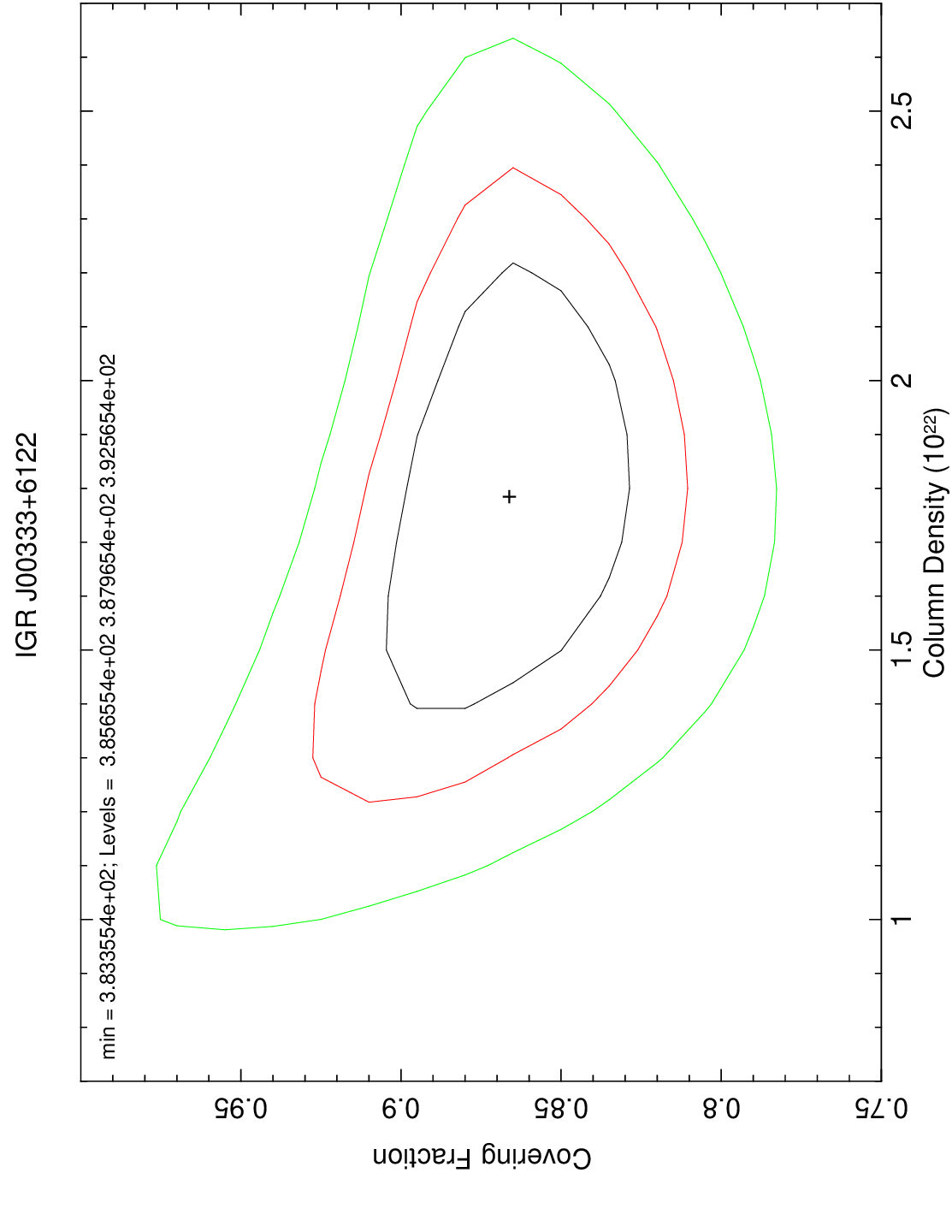}
\includegraphics[angle=-90, width=0.51\linewidth]{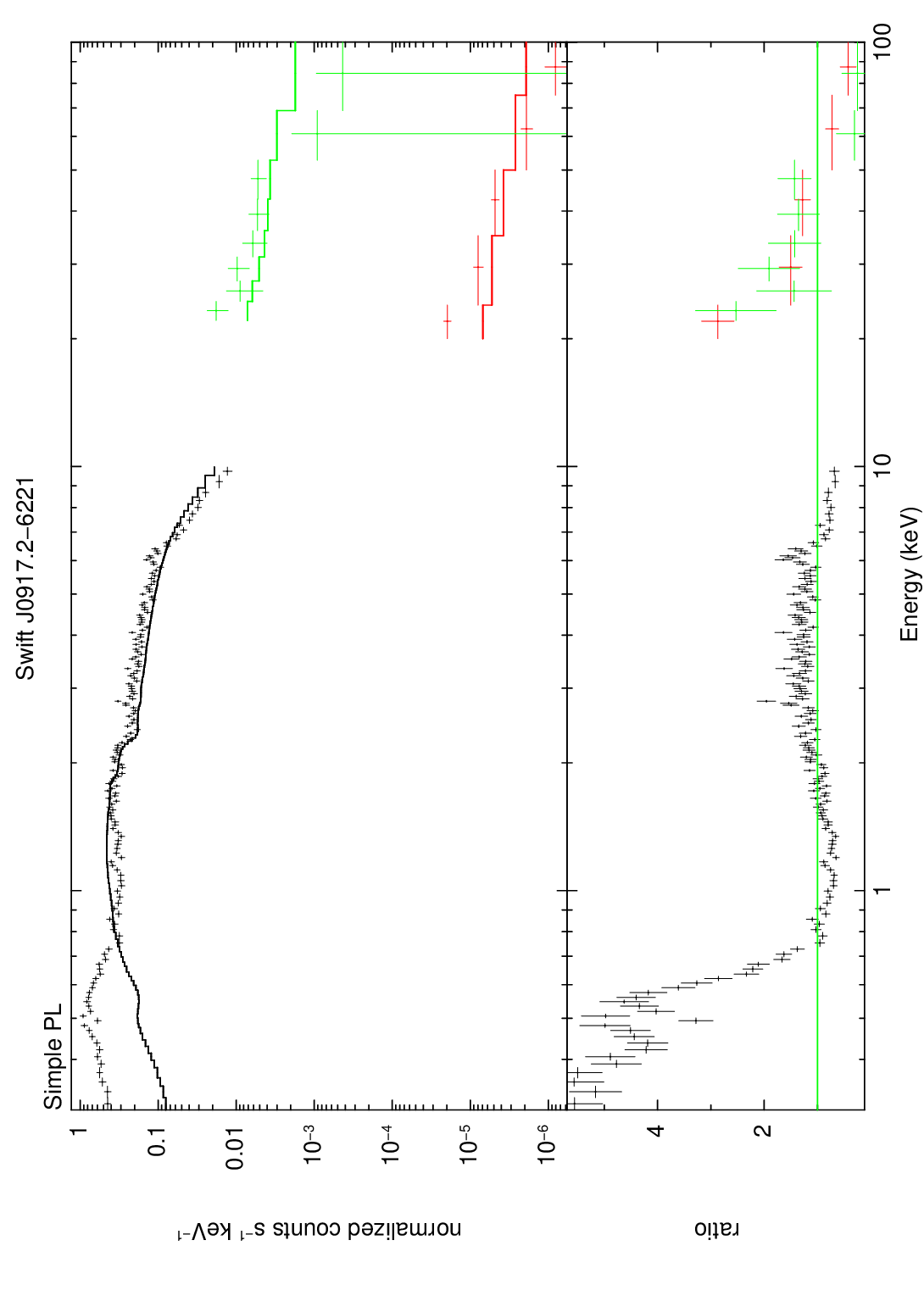}\\
\caption{\small{\emph{Left Panel}: confidence contour plot of the absorbing column density {\it vs.} 
covering fraction for IGR J00333+6122. 
\emph{Right Panel:} 0.5--10\,keV spectrum of Swift J0917.2-6221; the model employed is a simple power-
law absorbed by Galactic column density. The spectrum is very complex, showing a strong soft excess at 
low energies as well as absorption edges, the K$\alpha$/K$\beta$ iron line complex and a possible cut-
off at high energy.}}
\label{00333_swift}
\end{figure}
\end{small}

\begin{small}
\begin{figure}
\centering
\includegraphics[angle=-90, width=0.47\linewidth]{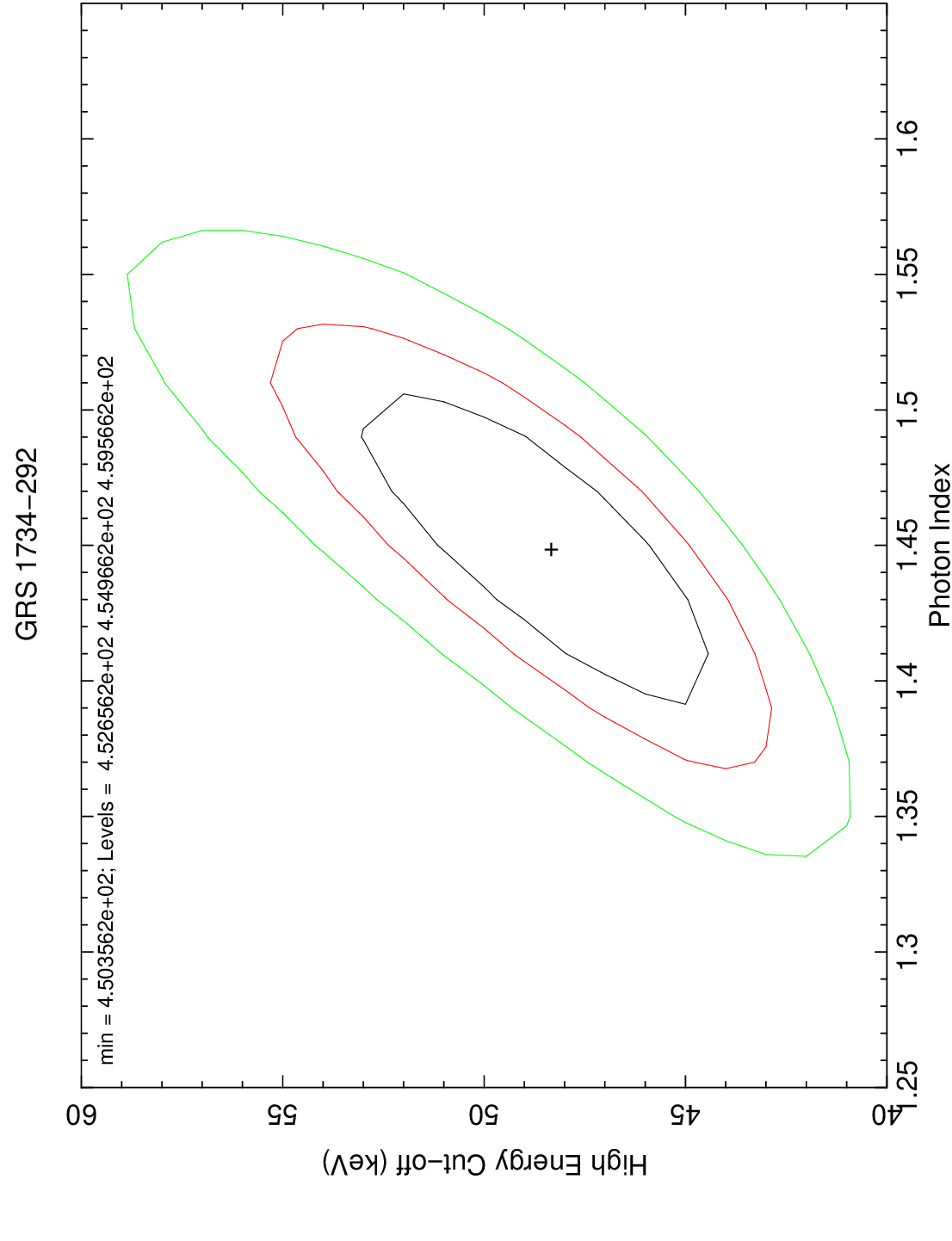}
\includegraphics[angle=-90, width=0.51\linewidth]{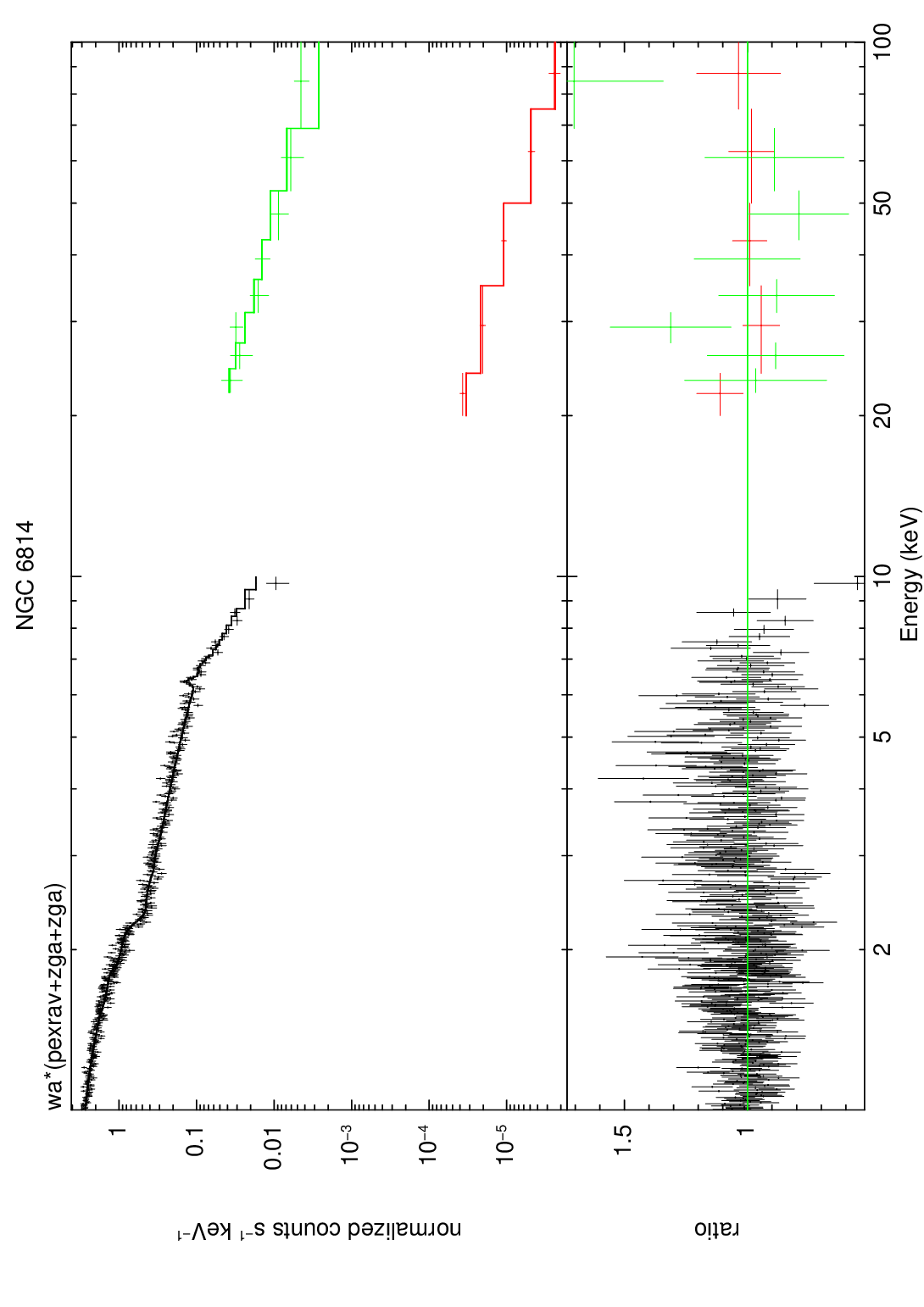}\\
\caption{\small{\emph{Left Panel}: confidence contour plot of the photon index {\it vs} 
the high energy cut-off for GRS 1734-292. Both parameters are very well constrained.
\emph{Right Panel:} Combined {\it XMM}, {\it Swift/BAT} and {\it INTEGRAL/IBIS} 
spectrum of NGC 6814. The model is an exponentially cut-off power-law, absorbed by Galactic 
N$_{\rm H}$, reflected from neutral material plus two gaussian components to model the Fe K$\alpha$ 
and K$\beta$ line complex.}}
\label{6814}
\end{figure}
\end{small}

{\it IGR J00333+6122.} The broad-band spectrum of this source is well described by a rather flat 
power-law ($\Gamma$=1.54$^{+0.36}_{-0.16}$) absorbed by a single layer of cold material 
(N$_{\rm H}$=1.79$^{+0.48}_{-0.43}$$\times$10$^{22}$ cm$^{-2}$)
covering 87$\pm$4\% of the central source, with both parameters very well constrained 
(see Figure~\ref{00333_swift}, left panel). The 0.5--110\,keV spectrum is also characterised 
by a high energy cut-off (although not well constrained)
located above $\sim$46\,keV and an upper limit on the reflection fraction of less than 3. 
The soft X-ray spectrum is instead characterised by the presence of a narrow iron line around 
6.4\,keV with an equivalent width of 43$^{+25}_{-22}$\,eV.

{\it Swift J0917.2-6221.} The broad-band spectrum of Swift J0917.2-6221 is very complex 
and not easily fitted with a simple model, such as the baseline one employed in our analysis. This is 
evident in Figure~\ref{00333_swift}, right panel, where the simple power-law fit to the data is shown. 
The model-to-data ratio puts in clear evidence the presence of low energy features, such as the soft 
excess and absorption edges, usually associated with O[VII] and O[VIII] lines. 
Around 6.4\,keV, the iron line complex is also present, with both K$\alpha$ and K$\beta$ lines detected. 
The data also show the presence of high energy cut-off at energies lower than 100\,keV.

{\it GRS 1734-292.} This source is known to have a high energy cut-off located at around 50\,keV (see 
\cite{molina09}). Indeed the {\it XMM/BAT/IBIS} broad-band spectrum is well-fitted by a 
cut-off power-law, with E$_{\rm cut}$=48$^{+7}_{-4}$\,keV and 
$\Gamma$=(see Figure~\ref{6814}, left panel). The continuum
slope is quite flat ($\Gamma$=1.45$^{+0.07}_{-0.06}$), absorbed by a column density
N$_{\rm H}$=(0.92$\pm$0.08)$\times$10$^{22}$\,cm$^{-2}$, while the 
reflection fraction is not well constrained, with an upper limit of about 0.34, consistent with the 
marginal detection of the iron line at around 6.4\,keV (EW$\le$46\,eV). 

{\it NGC 6814.} This source has been previously studied employing {\it ASCA} data 
(see \cite{molina06} and \cite{molina09}) and found to be a very variable AGN, as is also evident 
from the mismatch between the {\it ASCA} and {\it IBIS} data (yielding a cross-calibration 
constant of about 16; see Figure 2 in \cite{molina06}). Indeed, the 2--10\,keV flux measured by 
{\it ASCA} in May 1993 was 1.8$\times$10$^{-12}$erg\,cm$^{-2}$\,s$^{-1}$, while the 2--10\,keV flux 
measured by {\it XMM} in October 2009 was 2.75$\times$10$^{-11}$erg\,cm$^{-2}$\,s$^{-1}$. 
The {\it XMM/BAT/IBIS} broad-band spectrum (see Figure~\ref{6814}, right panel) is well fitted by a 
power-law ($\Gamma$=1.91$^{+0.03}_{-0.04}$) with no high energy cut-off, 
a complex iron line (a K$\alpha$ line with EW=151$^{+37}_{-34}$\,eV and a K$\beta$ line 
with EW=110$^{+40}_{-38}$\,eV) and a high reflection fraction (R=3.71$^{+0.04}_{-0.05}$), 
possibly due to the source variability rather than to a true physical 
property of NGC 6814.

\end{document}